\documentclass{sf2a-conf2011}
\usepackage{graphicx}
\usepackage{hyperref}
\usepackage[]{natbib}  
\usepackage{epstopdf}

\def\BibTeX{{\rm B\kern-.05em{\sc i\kern-.025em b}\kern-.08em
    T\kern-.1667em\lower.7ex\hbox{E}\kern-.125emX}}
\bibpunct{(}{)}{;}{a}{}{,}  %%%%%%%%%%%%%  A&A bibliography style
\newcommand{\degr}{{\mathrm{^\circ}}}
\begin{document}

\TitreGlobal{SF2A 2011}

%%-----------------------------------------------------------------
%%      the top matter
%%

\title{The Milky Way stellar populations in CFHTLS fields}

\runningtitle{The Milky Way stellar populations in CFHTLS fields}

\author{M. Guittet}\address{GEPI, Observatoire de Paris, CNRS, Universit\'e Paris Diderot; 5 Place Jules Janssen, 92190 Meudon, France}
\author{M. Haywood$^1$}
\author{M. Schultheis}\address{Observatoire de Besan\c con; 41 bis, avenue de l'Observatoire, 25000 Besan\c con, France}

%% Keep this line, even if the page will be settled afterwards.
\setcounter{page}{237}

%% To make the final index, repeat the authors here, in the format : Surname, Initial(s) 
\index{Guittet, M.}
\index{Haywood, M.}
\index{Schultheis, M.}

%%-----------------------------------------------------------------

\maketitle

%%-----------------------------------------------------------------
%%        The abstract
%% 
%%  Warning!  within the abstract:
%%  - do not use macros. 
%%  - do not use commands like: \cite, \citet, \citep ... etc.

\begin{abstract}
We investigate the characteristics of the thick disk in the  Canada -- France -- Hawaii -- Telescope
Legacy  Survey  (CFHTLS)  fields, complemented at bright magnitudes
with Sloan  Digital Sky  Survey (SDSS) data. The ([Fe/H], Z) 
distributions are derived in the W1 and W3 fields, and 
compared with simulated maps produced using the Besan\c con model.
It is shown that the thick disk,  represented in star-count models  
by a distinct component, is not an adequate description of the 
observed ([Fe/H], Z) distributions in these fields.
\end{abstract}

%% Insert the keywords (to appear in the ADS indexing)
%% Keywords must be separated by a comma

%%-----------------------------------------------------------------

\section{Introduction}
\label{intro}

Our knowledge of the characteristics of the thick disk remains limited in 
practically every aspects. Its structure on large scales ($>$kpc) is not well 
defined, either clumpy or smooth, and its connections with the collapsed
part of the halo or the old thin disk are essentially not understood. The spectrum of 
possible scenarios proposed to explain its formation is still very large 
and really discriminant constraints are rare. 
The  SDSS  photometric  survey has  provided  a wealth of new informations on
the thick disk, see in particular Ivezi\'c et al. (2008), Bond et al. (2010) and Lee et al. (2011).
However, the data have barely been directly confronted to star-count models, and 
little insights have been given on how the thick disk in these models really represents the 
survey data.
In the present work, we initiate such comparisons by comparing the Besan\c con model with metallicity and distance
information in the W1 and W3 CFHTLS fields, and provide a brief discussion of our results.

\section{Data description}
%%-------------------------

%\subsection{CFHTLS, SDSS}  
%%------------------------- 
Among the four fields  that make  the Wide  Survey, W1  and W3  cover  larger angular
surfaces (72 and 49 square degrees)  than W2 and W4 (both having 25 square degrees).  They point  towards higher latitudes  (--61.24$\degr$\ and
58.39$\degr$\ respectively) and are consequently less affected by dust
extinction,  and contain a  larger relative  proportion of  thick disk
stars. We will therefore focus on W1 and W3.  CFHTLS photometry starts
at a  substantially fainter magnitude  than the SDSS, missing  a large part of
the  thick disk. We  complemented the  CFHTLS catalogue  at the
bright  end  with  stars from  the  SDSS  not  present in  the  CFHTLS
fields. In  the final  catalogues, W1 contains  $\sim$ 139  000 stars,
with 16$\%$ from the SDSS, while  $\sim$ 132 000 stars are found in W3
field, with  31$\%$ coming from  the Sloan.\\ W1  and W3 are  at large
distances above the galactic plane.  The dust extinction is very small
at these  latitudes.  For example the  Schlegel map \citep{schlegel98}
estimates  for   W1  an  absorption   coefficient  Av  of   0.087  while
\citet{Jones11}    give   Av=0.113.    The   extinction    models   of
\citet{Arenou92} or  \citet{Hakkila97} estimate  Av values to  0.1 and
0.054  respectively. We briefly discuss the effect of extinction on
distance determination and metallicities in 4.1.

\section{Comparisons between the Besan\c con model and CFHTLS/SDSS data: Hess diagrams}
%%---------------------

\subsection{The  Besan\c con  model} Simulations  were made  using the
Besan\c     con     model     (\citet{Robin03},     \citet{Haywood97},
\citet{Bienayme87})   online   version.   The  model   includes   four
populations: the bulge, the thin disk, the thick disk and the halo. 
The metallicities of the thick disk and the halo in the online version
of the model  (--0.78 and --1.78 dex respectively) were shifted (to --0.6 dex and -1.5 dex)
to comply with more generally accepted values, and in particular with 
values derived from the Sloan data (see \citet{Lee11}, who shows that
the thick  disk have a metallicity  [Fe/H] = --0.6  dex roughly
independant  of vertical  distances, and (\citet{ivezic08},
\citet{Bond10}, \citet{Sesar11}, \citet{Carollo10} or \citet{Dejong10}
for the  inner halo metallicity, estimated to be about --1.5  dex). 
The thick disk has a scale height of 800 pc and a local stellar density $\rho_0$
of 6.8  $\%$ of the local thin  disk, while the stellar halo is described by
a power law with a flattening and a local density of 0.6\%.
Simulations where made assuming photometric errors as described in the SDSS.

\subsection{Hess diagrams}
\begin{figure}[ht!]
\centering
\includegraphics[width=0.35\textwidth,clip,angle=-90]{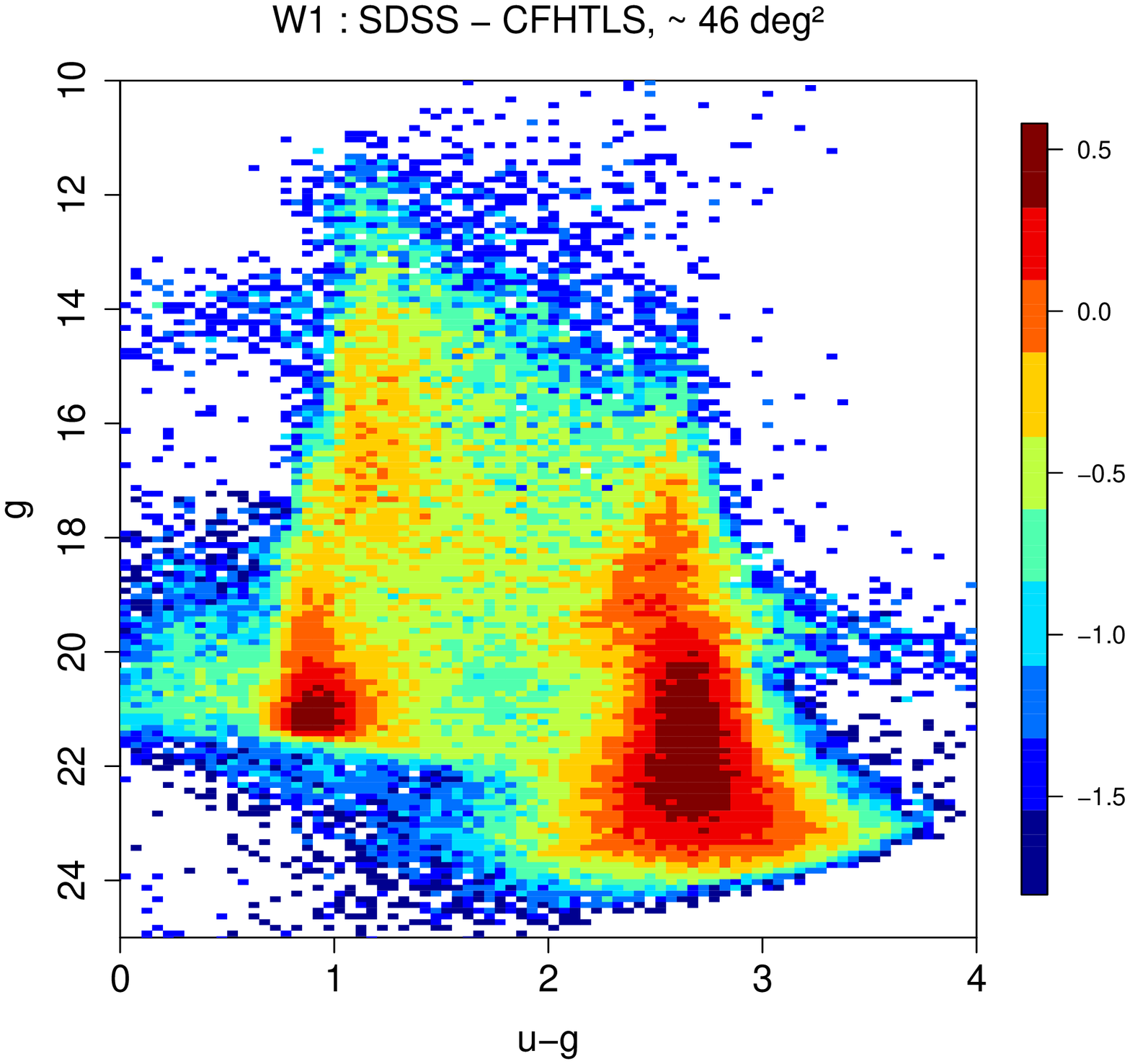}
\includegraphics[width=0.35\textwidth,clip,angle=-90]{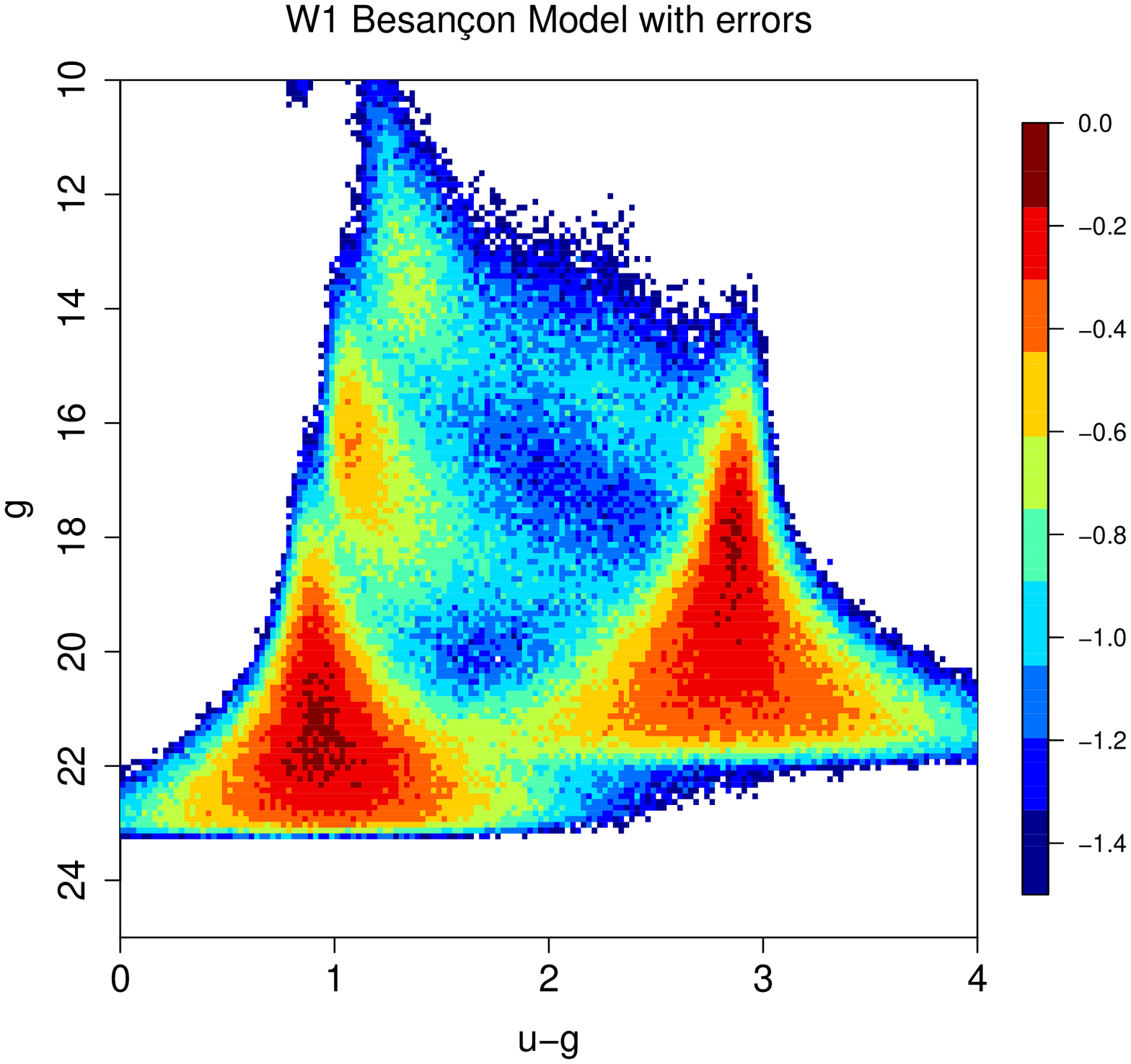}
 \caption{{\bf Left:} CFHTLS/SDSS u--g Hess diagram in W1 field. {\bf Right:} The same as the left graphic but for the Besan\c con model stars to which observationnal errors from SDSS data have been added. The successive ridge lines due to the thin disk, thick disk and the halo main sequence stars are clearly distinguishable between 0.7$<$u-g$<$1.5.}
  \label{fig1}
\end{figure}

The distributions of CFHTLS/SDSS and model stars in the g versus u--g color magnitude diagram (CMD) are shown in Fig.~\ref{fig1}.
For both diagrams, faint blue stars (u--g $\sim$ 0.9, g$>$18) are clearly discernible and correspond to the galactic halo. 
The concentration of stars at g$<$18, u--g $\sim$1.1, corresponds to disk stars and in particular thick disk stars. 
Because of the SDSS saturation at g=14 which does not allow to have a representative sample of thin disk stars, our data sample 
is mainly composed of thick disk and halo stars. The Besan\c con model shows a distinct separation between
thin disk stars (u--g$\sim$1.3, g$<$14-15) and thick disk stars (u--g$\sim$1.1, 15$<$g$<$18) which cannot be check with the present 
data.\\

\section{Comparisons between the Besan\c con model and CFHTLS/SDSS data: ([Fe/H], Z) distributions}

\subsection{Metallicity and photometric distance determinations}
\citet{Juric08} and \citet{ivezic08} have published calibrations of the metallicity and photometric parallax as a function of ugri magnitudes.
The metallicity calibration has been revised in \citet{Bond10} :
\begin{eqnarray} 
\label{feh}  \mathrm{[Fe/H]} & = & \mathrm{A + Bx + Cy + Dxy + Ex^2 + Fy^2 + Gx^2y + Hxy^2 + Ix^3 + Jy^3 }
\end{eqnarray} 

where x  = $u\--g$, y = $g\--r$  and (A--J) = (--13.13,
14.09, 28.04,  --5.51, --5.90, --58.58, 9.14,  --20.61, 0.0, 58.20).\\
This  relation  has  been  determined   for  F  and  G  stars  and  is
consequently applicable in the range :  0.2 $< g\--r <$ 0.6 and --0.25
+ 0.5($u\--g$) $< g\--r <$ 0.05 + 0.5($u\--g$).  This calibration only
extends  to --0.2  dex.   Observed vertical  distances  $Z$ have  been
calculated using $\normalsize{ Z\ =  \ \mathrm{D \ sin(b)} }$, b being
the  latitude  of  the   star.  Photometric  distances  $D$,  such  as
$\normalsize{m_{r}  \-- M_{r}  = 5\  log(D) \--  5}$,  were determined
using  the absolute  magnitude calibration  of  \citet{ivezic08} which
depends on  the metallicity  and on $g-i$  colours.\\ 

For the  highest extinction values  given by
\citet{Jones11}, the impact on metallicities, as can be estimated using  Eq.~\ref{feh}
and the absolute  magnitude   relation  of  \citet{ivezic08} are at most  of 0.15 dex
near g--r=0.5 at  solar metallicities and 0.1 dex  at [Fe/H]= --1 dex.
Distances will  be affected  at most by about  20\%  at solar
metallicities and 15\% at [Fe/H]= --1 dex at g--r near 0.40-0.45.\\

\subsection{ ([Fe/H], Z) distributions}

\begin{figure}[ht!]
\centering
\includegraphics[width=0.35\textwidth,clip,angle=-90]{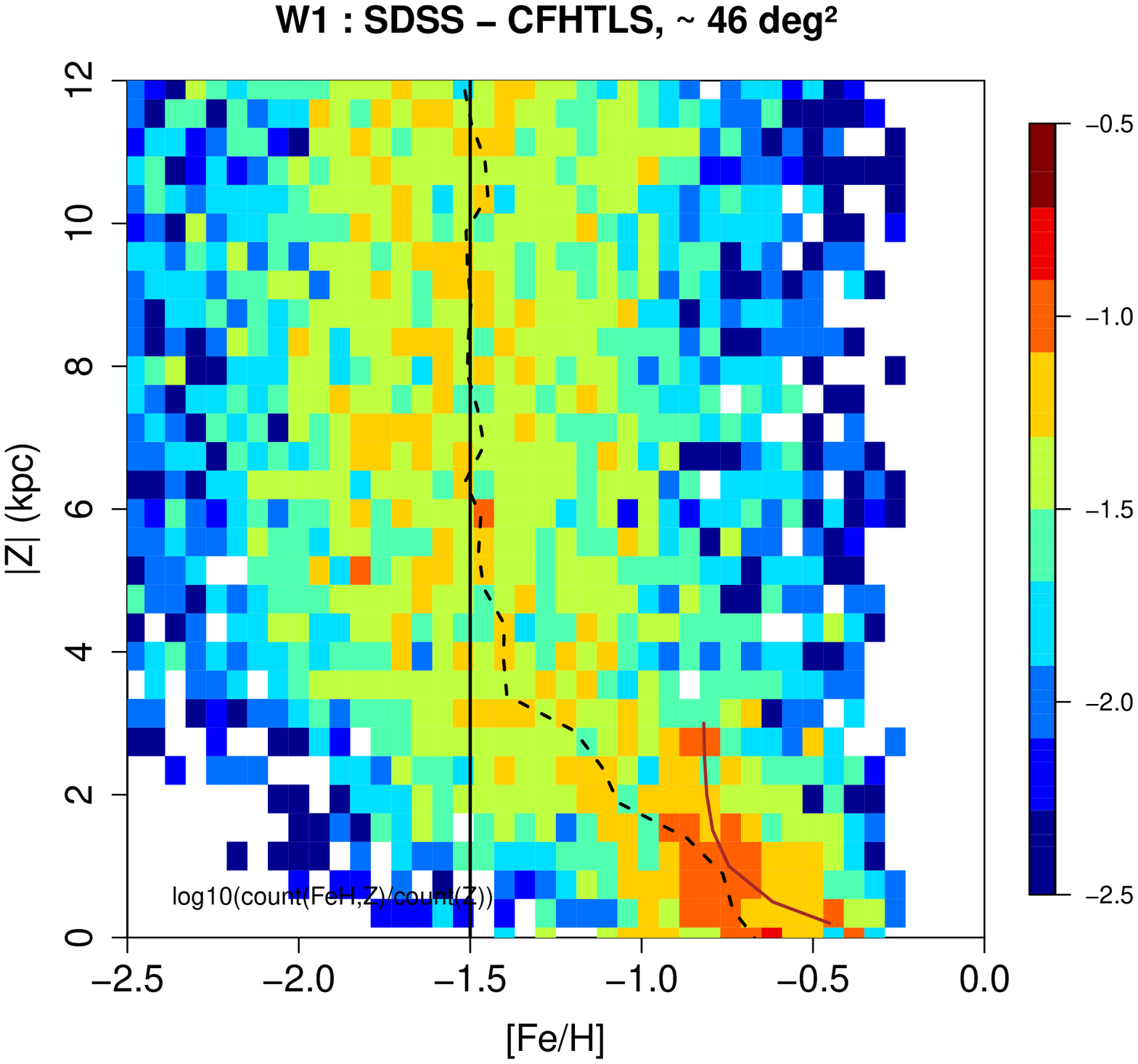}
\includegraphics[width=0.35\textwidth,clip,angle=-90]{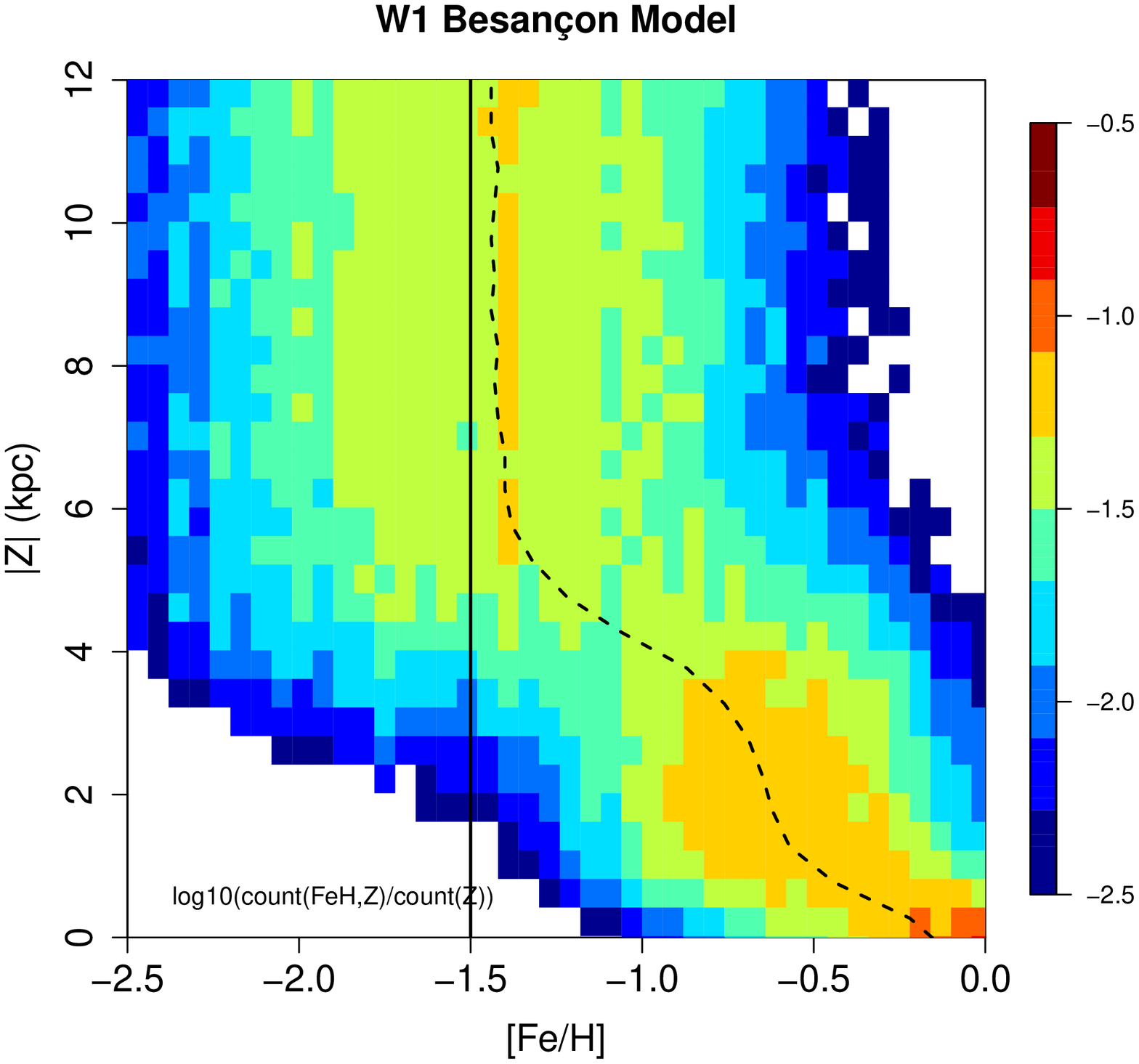}
 \caption{{\bf Left:} ([Fe/H], Z) distribution for CFHTLS/SDSS data in W1 field. {\bf Right:} Besan\c con model ([Fe/H], Z) distribution in W1 field. The continuous line on the left plot shows the median metallicity formula (A2) derived by \citet{Bond10}. The dotted line represents the median metallicity per bin of 0.5 kpc.  }
  \label{fig2}
\end{figure}

We generated catalogues with the model in the direction of W1 and W3, deriving the Z height above the plane from
simulated distances and metallicities from the assumed metallicity distributions of each population.
In Fig.~\ref{fig2}  we present ([Fe/H], Z) distributions  for both the
data and the model. The dotted  line is the median metallicity per bin
of 0.5  kpc.  The continuous line  is the median  metallicity for disk
stars as shown  by \citet{Bond10} and follows rather  well the disk distribution in
our  data.  We find  similar  results  as  \citet{Bond10} :  the  halo
dominates the star counts above 3 kpc with a mean metallicity of about
--1.5  dex. \citet{Sesar11} studied  the four  CFHTLS Wide  fields but
with magnitudes corrected for ISM extinction. They found the mean halo
metallicity in the range between --1.4  and --1.6 dex. Our estimate of the
extinction   effect would shift metallicities to about  0.15 dex at most, and 
shows  that  our  mean  halo  metallicity  is  in  good
agreement with  their estimates.   

The interesting point worth of notice is the conspicuous, distinct, pattern that
represents the thick disk in the model and which clearly is absent in the data.
As expected, the standard thick disk model dominates the counts between 1 and 4 kpc, 
while in the data, the thick disk seems to be less extended, and does not appear
as a distinct component between the thin disk and the halo. The vertical resolution 
of the observed distribution prevents any clear statement concerning the transition
from the thin to thick disk, although it is apparent that the model is 
at variance with the data. 
This result raises the interesting question of the connections (or lack of) between the 
thin and thick disks. Almost since its discovery, it has been suggested that the thick disk
is more akin to an extended thin disk \citep{Norris87}. Our knowledge of the thick disk
more than twenty years later does not permit us to draw any firm conclusion on that point.

\section{Conclusion}
%%-------------------
Investigation of the ([Fe/H], Z) distribution in the CFHTLS Wide fields  does not 
seem to show a thick disc component as prominent and distinct as predicted by  
standard star-count models. The mean halo metallicity found to be --1.5 dex is in agreement with previous studies (e.g \citet{Bond10}, \citet{Sesar11}).
 The behavior of models must be studied on more extensive 
data sets in order to assess the necessary adjustments and to better characterise
the thick disk.

\bibliographystyle{aa}  % A&A bibliography style file (aa.bst)
\bibliography{guittet} % your references in file: Yourfile.bib

\begin{thebibliography}{15}
\expandafter\ifx\csname natexlab\endcsname\relax\def\natexlab#1{#1}\fi

\bibitem[{{Arenou} {et~al.}(1992){Arenou}, {Grenon}, \& {Gomez}}]{Arenou92}
{Arenou}, F., {Grenon}, M., \& {Gomez}, A. 1992, \aap, 258, 104

\bibitem[{{Bienayme} {et~al.}(1987){Bienayme}, {Robin}, \&
  {Creze}}]{Bienayme87}
{Bienayme}, O., {Robin}, A.~C., \& {Creze}, M. 1987, \aap, 180, 94

\bibitem[{{Bond} {et~al.}(2010){Bond}, {Ivezi{\'c}}, {Sesar}, {Juri{\'c}},
  {Munn}, {Kowalski}, {Loebman}, {Ro{\v s}kar}, {Beers}, {Dalcanton},
  {Rockosi}, {Yanny}, {Newberg}, {Allende Prieto}, {Wilhelm}, {Lee},
  {Sivarani}, {Majewski}, {Norris}, {Bailer-Jones}, {Re Fiorentin}, {Schlegel},
  {Uomoto}, {Lupton}, {Knapp}, {Gunn}, {Covey}, {Allyn Smith}, {Miknaitis},
  {Doi}, {Tanaka}, {Fukugita}, {Kent}, {Finkbeiner}, {Quinn}, {Hawley},
  {Anderson}, {Kiuchi}, {Chen}, {Bushong}, {Sohi}, {Haggard}, {Kimball},
  {McGurk}, {Barentine}, {Brewington}, {Harvanek}, {Kleinman}, {Krzesinski},
  {Long}, {Nitta}, {Snedden}, {Lee}, {Pier}, {Harris}, {Brinkmann}, \&
  {Schneider}}]{Bond10}
{Bond}, N.~A., {Ivezi{\'c}}, {\v Z}., {Sesar}, B., {et~al.} 2010, \apj, 716, 1

\bibitem[{{Carollo} {et~al.}(2010){Carollo}, {Beers}, {Chiba}, {Norris},
  {Freeman}, {Lee}, {Ivezi{\'c}}, {Rockosi}, \& {Yanny}}]{Carollo10}
{Carollo}, D., {Beers}, T.~C., {Chiba}, M., {et~al.} 2010, \apj, 712, 692

\bibitem[{{de Jong} {et~al.}(2010){de Jong}, {Yanny}, {Rix}, {Dolphin},
  {Martin}, \& {Beers}}]{Dejong10}
{de Jong}, J.~T.~A., {Yanny}, B., {Rix}, H.-W., {et~al.} 2010, \apj, 714, 663

\bibitem[{{Hakkila} {et~al.}(1997){Hakkila}, {Myers}, {Stidham}, \&
  {Hartmann}}]{Hakkila97}
{Hakkila}, J., {Myers}, J.~M., {Stidham}, B.~J., \& {Hartmann}, D.~H. 1997,
  \aj, 114, 2043

\bibitem[{{Haywood} {et~al.}(1997){Haywood}, {Robin}, \& {Creze}}]{Haywood97}
{Haywood}, M., {Robin}, A.~C., \& {Creze}, M. 1997, \aap, 320, 428

\bibitem[{{Ivezi{\'c}} {et~al.}(2008){Ivezi{\'c}}, {Sesar}, {Juri{\'c}},
  {Bond}, {Dalcanton}, {Rockosi}, {Yanny}, {Newberg}, {Beers}, {Allende
  Prieto}, {Wilhelm}, {Lee}, {Sivarani}, {Norris}, {Bailer-Jones}, {Re
  Fiorentin}, {Schlegel}, {Uomoto}, {Lupton}, {Knapp}, {Gunn}, {Covey},
  {Smith}, {Miknaitis}, {Doi}, {Tanaka}, {Fukugita}, {Kent}, {Finkbeiner},
  {Munn}, {Pier}, {Quinn}, {Hawley}, {Anderson}, {Kiuchi}, {Chen}, {Bushong},
  {Sohi}, {Haggard}, {Kimball}, {Barentine}, {Brewington}, {Harvanek},
  {Kleinman}, {Krzesinski}, {Long}, {Nitta}, {Snedden}, {Lee}, {Harris},
  {Brinkmann}, {Schneider}, \& {York}}]{ivezic08}
{Ivezi{\'c}}, {\v Z}., {Sesar}, B., {Juri{\'c}}, M., {et~al.} 2008, \apj, 684,
  287

\bibitem[{{Jones} {et~al.}(2011){Jones}, {West}, \& {Foster}}]{Jones11}
{Jones}, D.~O., {West}, A.~A., \& {Foster}, J.~B. 2011, \aj, 142, 44

\bibitem[{{Juri{\'c}} {et~al.}(2008){Juri{\'c}}, {Ivezi{\'c}}, {Brooks},
  {Lupton}, {Schlegel}, {Finkbeiner}, {Padmanabhan}, {Bond}, {Sesar},
  {Rockosi}, {Knapp}, {Gunn}, {Sumi}, {Schneider}, {Barentine}, {Brewington},
  {Brinkmann}, {Fukugita}, {Harvanek}, {Kleinman}, {Krzesinski}, {Long},
  {Neilsen}, {Nitta}, {Snedden}, \& {York}}]{Juric08}
{Juri{\'c}}, M., {Ivezi{\'c}}, {\v Z}., {Brooks}, A., {et~al.} 2008, \apj, 673,
  864

\bibitem[{{Lee} {et~al.}(2011){Lee}, {Beers}, {An}, {Ivezi{\'c}}, {Just},
  {Rockosi}, {Morrison}, {Johnson}, {Sch{\"o}nrich}, {Bird}, {Yanny},
  {Harding}, \& {Rocha-Pinto}}]{Lee11}
{Lee}, Y.~S., {Beers}, T.~C., {An}, D., {et~al.} 2011, \apj, 738, 187

\bibitem[{{Norris}(1987)}]{Norris87}
{Norris}, J. 1987, \apjl, 314, L39

\bibitem[{{Robin} {et~al.}(2003){Robin}, {Reyl{\'e}}, {Derri{\`e}re}, \&
  {Picaud}}]{Robin03}
{Robin}, A.~C., {Reyl{\'e}}, C., {Derri{\`e}re}, S., \& {Picaud}, S. 2003,
  {\aa}p, 409, 523

\bibitem[{{Schlegel} {et~al.}(1998){Schlegel}, {Finkbeiner}, \&
  {Davis}}]{schlegel98}
{Schlegel}, D.~J., {Finkbeiner}, D.~P., \& {Davis}, M. 1998, \apj, 500, 525

\bibitem[{{Sesar} {et~al.}(2011){Sesar}, {Juri{\'c}}, \&
  {Ivezi{\'c}}}]{Sesar11}
{Sesar}, B., {Juri{\'c}}, M., \& {Ivezi{\'c}}, {\v Z}. 2011, \apj, 731, 4

\end{thebibliography}

\end{document}